\documentclass{PoS}

\usepackage{amsmath}
\usepackage{dcolumn}
\usepackage{multirow}
\newcolumntype{d}{D{.}{.}{1.2}}

\title{$D$-meson semileptonic form factors at zero momentum transfer in (2+1+1)-flavor lattice QCD}

\ShortTitle{$D$-meson semileptonic form factors at zero momentum transfer}

\author{\speaker{Thomas Primer}, {Doug Toussaint}
         \\
        Physics Department, University of Arizona, Tucson, AZ 85721, USA\\
        E-mail: \email{thomas.primer@gmail.com}}

\author{Claude Bernard, Javad Komijani
\thanks{Present address: Institute for Advanced Study, Technische Univerit\"at M\"unchen, 85748 Garching, Germany}\\
  Department of Physics, Washington University, St. Louis, MO 63130, USA
  }
\author{Carleton DeTar\\
  Department of Physics and Astronomy, University of Utah, Salt Lake City, UT 84112, USA
  }
\author{Aida El-Khadra\\
  Department of Physics, University of Illinois, Urbana, IL 61801, USA
  }
\author{Elvira G\'amiz\\
  CAFPE and Departamento de F\'{\i}sica Te\'orica y del del Cosmos, Universidad de Granada, E-18071 Granada, Spain
  }
\author{Andreas Kronfeld
\thanks{Also at Institute for Advanced Study, Technische Univerit\"at M\"unchen, 85748 Garching, Germany}
, James Simone, Ruth S. Van de Water\\
Fermi National Accelerator Laboratory, Batavia, IL 60510, USA
  }
\author{Fermilab Lattice and MILC Collaborations}

\abstract{We present a calculation of the $D\to K \ell \nu$ and $D\to\pi \ell \nu$ semileptonic form factors at $q^2=0$, which enable determinations of the CKM matrix elements $\lvert{V_{cs}}\rvert$ and $\lvert{V_{cd}}\rvert$, respectively.
We use gauge-field configurations generated by the MILC collaboration with four flavors of highly-improved staggered (HISQ) quarks, analyzing several ensembles including those with physical pion masses and approximate lattice spacings ranging from 0.12~fm to 0.042~fm. We also use the HISQ action for the valence quarks.
We employ twisted boundary conditions to calculate the form factors at zero momentum transfer directly. We use 
heavy-light-meson chiral perturbation theory modified for energetic pions and kaons, and supplemented by terms to describe the lattice-spacing dependence, to obtain preliminary results at the physical point and in the continuum limit.}

\FullConference{The 33rd International Symposium on Lattice Field Theory\\
                 14 -18 July  2015\\
                 Kobe International Conference Center, Kobe, Japan}

\begin{document}

\section{Motivation}

The unitarity of the Standard-Model Cabibbo-Kobayashi-Maskawa (CKM) matrix leads to the relationship between elements of the second row $\lvert{V_{cd}}\rvert^2+\lvert{V_{cs}}\rvert^2+\lvert{V_{cb}}\rvert^2=1$.
The elements $\lvert{V_{cd}}\rvert$ and $\lvert{V_{cs}}\rvert$  can be obtained from leptonic $D$- and $D_s$-meson decays by combining experimental rate measurements with lattice-QCD calculations of the decay constants.
The Fermilab Lattice and MILC Collaborations recently calculated $f_{D}$ and $f_{D_s}$ to high precision~\cite{FNALMILC_DC} and the CKM elements are limited by experimental uncertainties:
\begin{equation}  \lvert{V_{cd}}\rvert = 0.217(1)_{\rm{LQCD}}(5)_{\rm{expt}}(1)_{\rm{EM}}\ ,
 \ \ \ \ \ \ \  \lvert{V_{cs}}\rvert = 1.010(5)_{\rm{LQCD}}(18)_{\rm{expt}}(6)_{\rm{EM}}. \end{equation}

These elements can also be obtained via the semileptonic decays $D\to K \ell \nu$ and $D\to \pi \ell \nu$, which require lattice-QCD calculations of the form factors $f_+(q^2)$.
Combining the currently most precise lattice results for $f_+(0)$ from the HPQCD Collaboration~\cite{HPQCD_1,HPQCD_2}
with the corresponding experimental averages from the Heavy Flavor Averaging Group~\cite{HFAG} yields,
\begin{equation} \lvert{V_{cd}}\rvert = 0.214(9)_{\rm{LQCD}}(3)_{\rm{expt}}\ ,
  \ \ \ \ \ \  \lvert{V_{cs}}\rvert = 0.977(14)_{\rm{LQCD}}(7)_{\rm{expt}} \,, \end{equation}
where the errors from lattice QCD are two to three times larger than from experiment.
Our goal is to bring the lattice-QCD errors on the zero-momentum-transfer form factors $f_+(0)$ to a level at or below the experimental uncertainties in $f_+(0) \times \lvert{V_{cd(s)}}\rvert$, so that lattice QCD is no longer the limiting source of uncertainty in determining $\lvert{V_{cs}}\rvert$ and $\lvert{V_{cd}}\rvert$ via semileptonic decays.
Such precision may shed light on the slight tension with second-row unitarity seen in leptonic decays~\cite{Rosner:2015wva}.

\section{Method}

The vector form factor $f_+(0)$ can be determined via the hadronic matrix elements for the flavor changing vector current, $V^\mu = \bar{q}\gamma^\mu c$.
Because the local lattice vector current is not conserved, this approach requires the calculation of a renormalization factor to extract the physical form factor. 
Instead we follow the approach introduced by HPQCD in Ref.~\cite{HPQCD_scalar}, in which the Ward Identity is used to obtain $f_0(q^2)$ from the matrix element of the scalar current, $S = \bar{q}c$, with absolute normalization:
\begin{eqnarray}
  \langle K(\pi) \lvert{S}\rvert{D}\rangle &=& \frac{M^2_D - M^2_{K(\pi)}}{m_c-m_{s(d)}}f_0^{D\to K(\pi)}(q^2),
\end{eqnarray}
We then exploit the kinematic constraint $f_+(0) = f_0(0)$ to arrive at an absolutely normalized value for $f_+(0)$.

\begin{figure}
  {\centering \vspace{-18pt}
    ${\vcenter{\hbox{\includegraphics[width=0.4\linewidth]{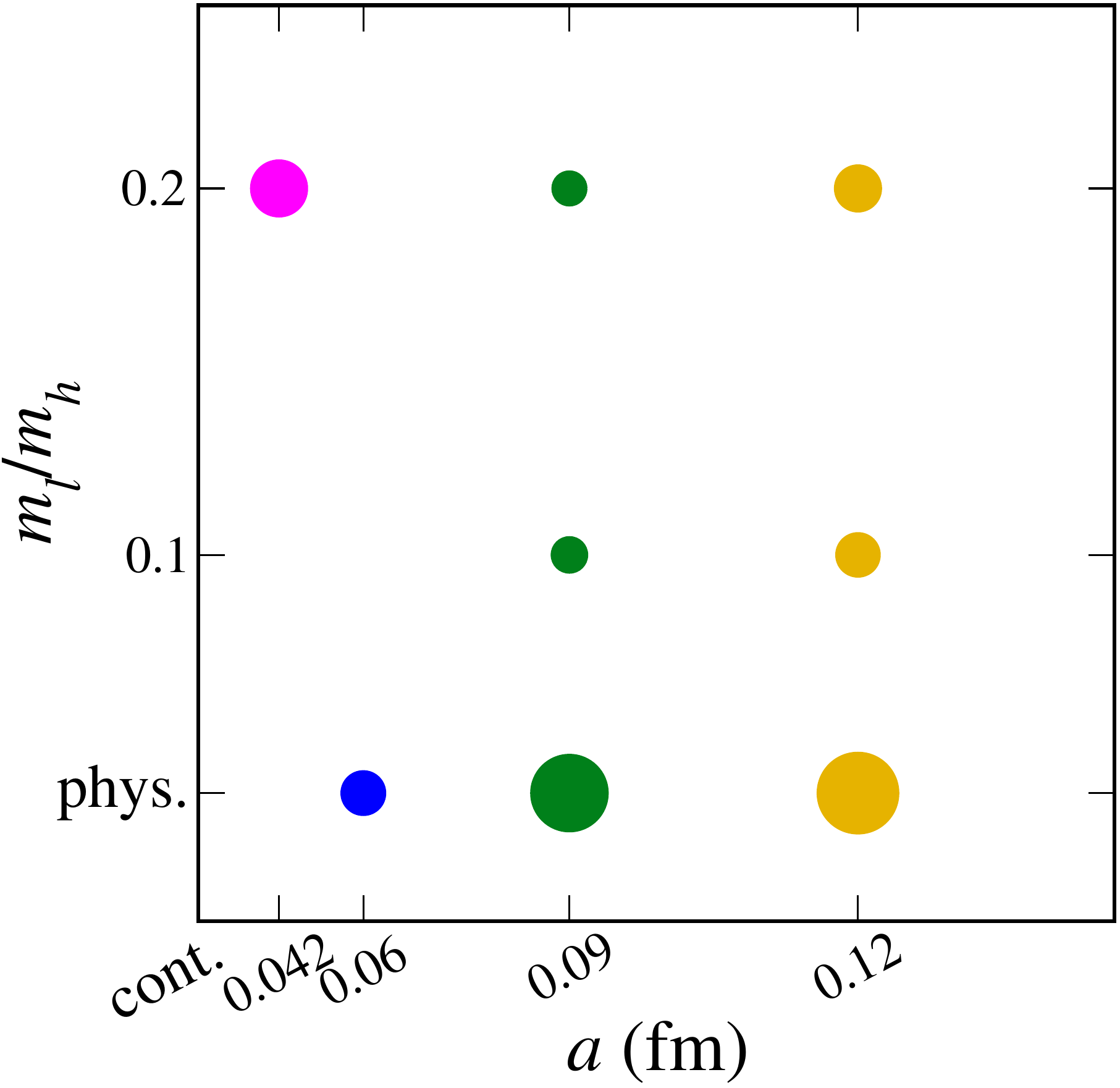}}}}$
    ${\vcenter{\hbox{\includegraphics[width=0.5\linewidth]{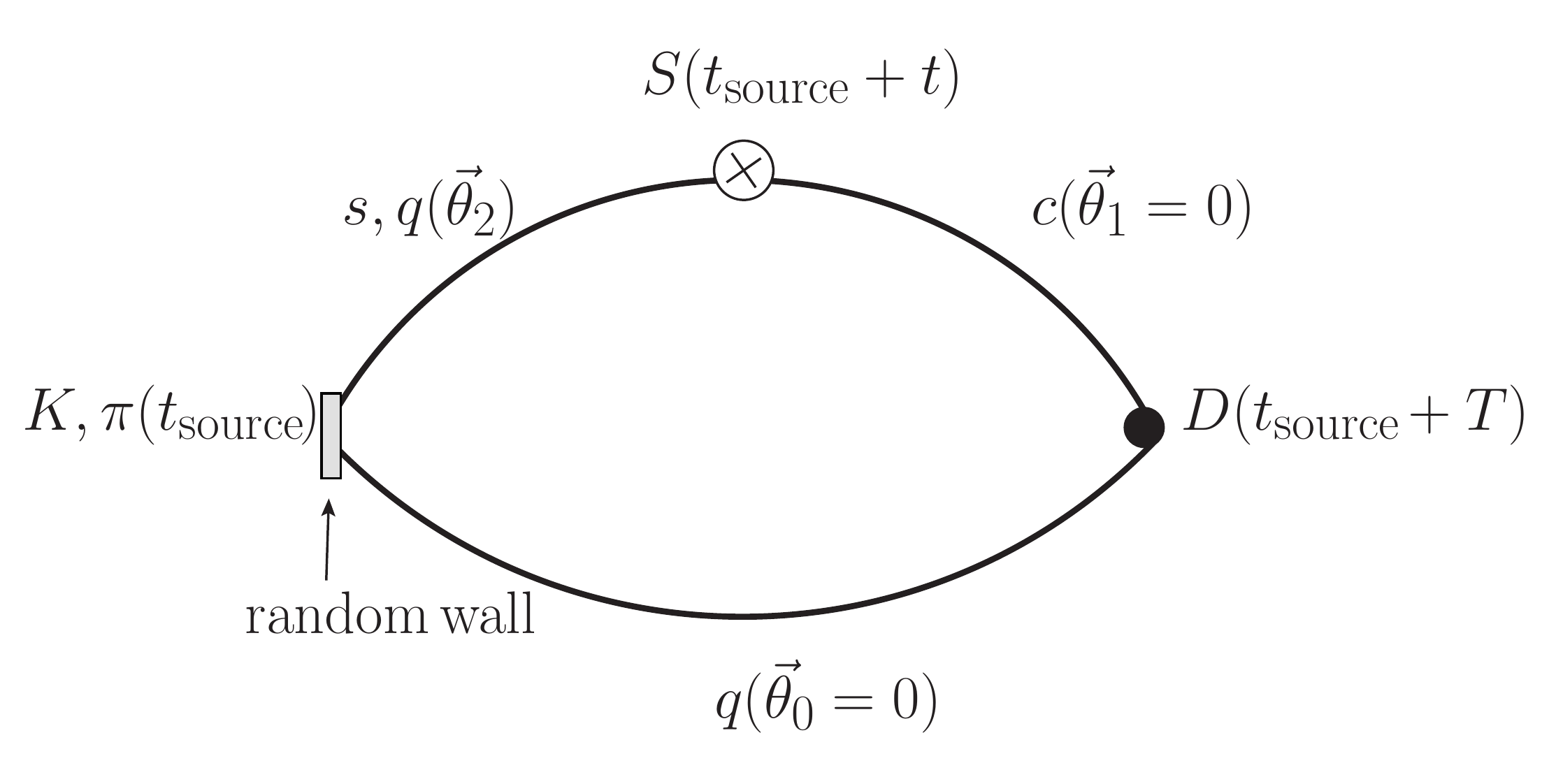}}}}$ \\
    \caption{(left) Ensembles used in this calculation, where the symbol area is proportional to the number of configurations times the number of sources. 
      (right) Structure of three-point correlators. We employ a random wall source at $t_{\rm{source}}$ and an extended source created from the spectator quark at $t_{\rm{source}}+T$.
      The $\theta_i$ denote the twist on each propagator, which is non-zero only for $\theta_2$ on the recoil propagator.
      The scalar current $S$ is inserted at $t_{\rm{source}}+t$. }\label{ensembles}
  }
\end{figure}

This calculation uses the MILC (2+1+1)-flavor HISQ ensembles \cite{Bazavov:2010ru} listed in Fig.~\ref{ensembles}.
The light, strange, and charm valence quarks are also simulated with the HISQ action \cite{Follana:2006rc,milc_hisq}.
The ensemble set includes three physical quark mass ensembles and lattice spacings
down to 0.042 fm.
Each ensemble has $N_{\rm{cfg}}\times n_{\rm{tsrc}}$ of at least 3000 and an $M_\pi L > 3.5$.

We perform this calculation directly at $q^2=0$ by using twisted boundary conditions to tune the momenta
of the child particles.
A twist of $\theta_2$ (see Fig.~\ref{ensembles}) in each spatial direction gives a momentum $\vec{p} = \theta_2 \pi(1,1,1)/L$ to the $K(\pi)$.
Due to the large mass difference between the $D$ meson and daughter meson, we need large momenta with $\theta_2$ in the range 2 to 5.
The calculation requires three-point $D\to K$ and $D\to\pi$ correlators with the structure shown in Fig.~\ref{ensembles}.
It also requires two-point $D$, $K$ and $\pi$ correlators, with versions for both twisted and zero momentum kaons and pions.

\section{Correlator analysis}

\begin{figure}
  {\centering \vspace{-18pt}
    \includegraphics[width=0.42\linewidth]{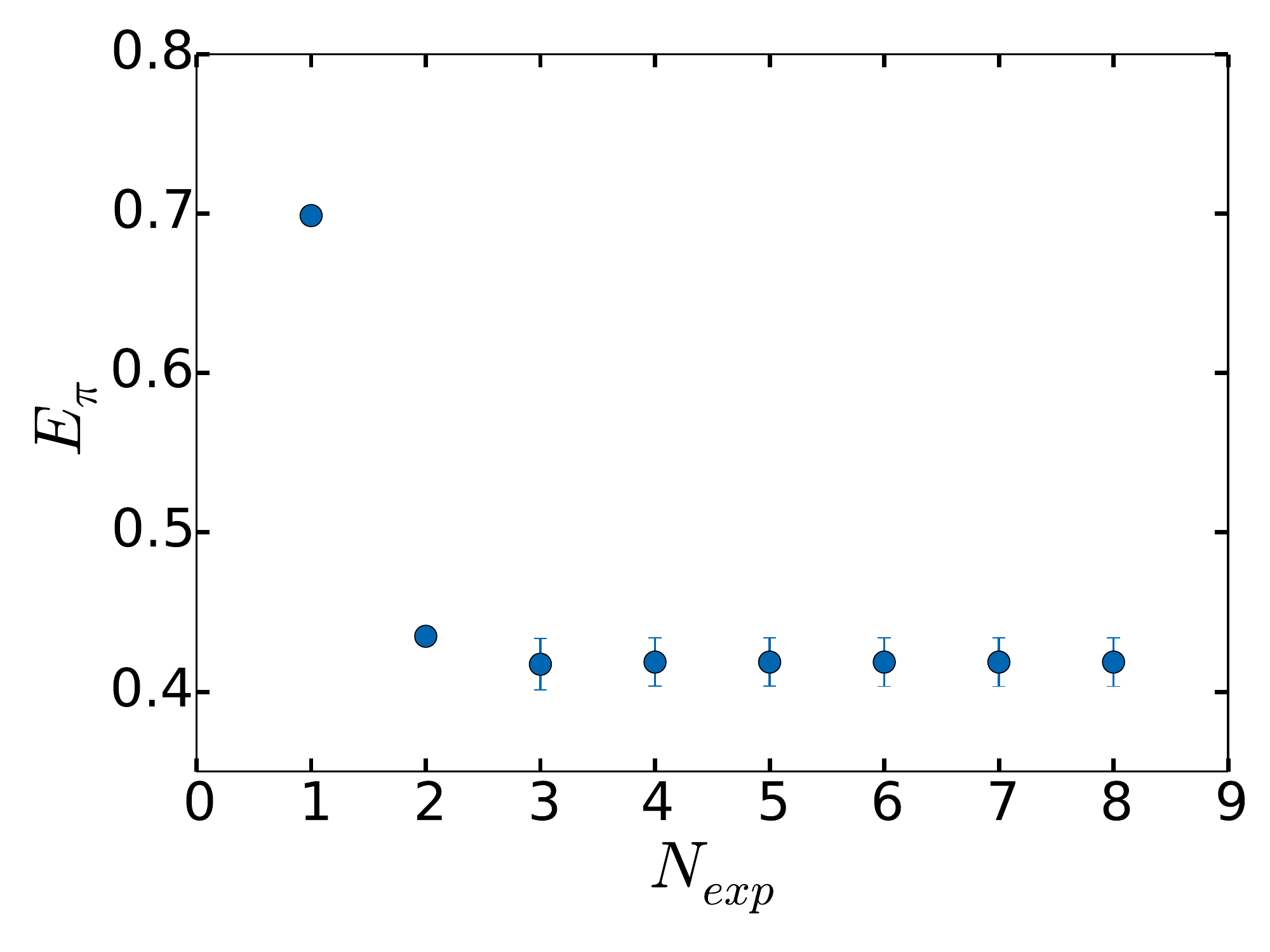}
    \includegraphics[width=0.42\linewidth]{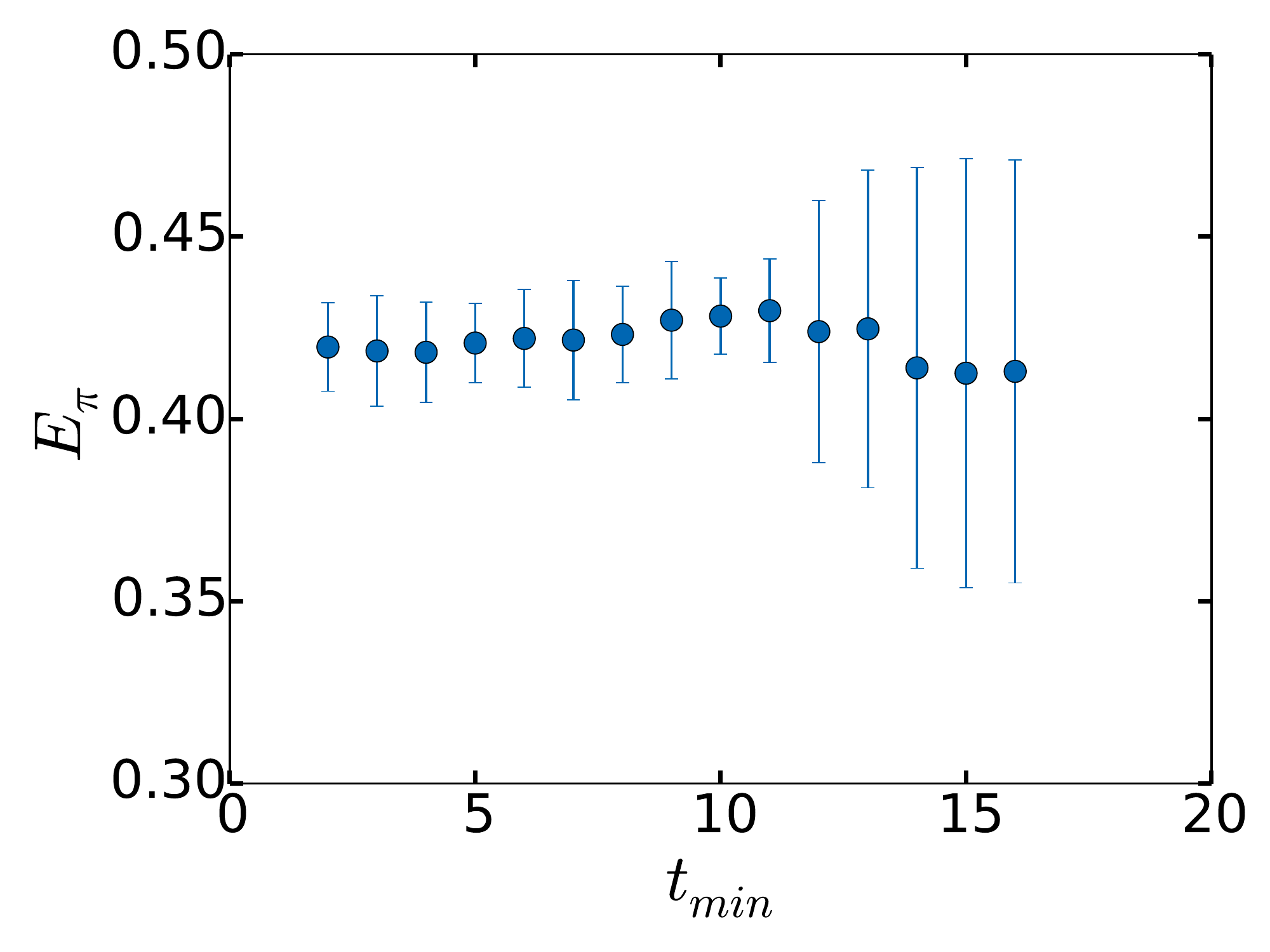} \\
    \caption{Pion energy values from the 0.09 fm physical quark mass ensemble as a function of the number of states $N_{\rm{exp}}$ with $t_{\rm{min}}$ fixed at 4 (left)
      and as a function of $t_{\rm{min}}$ with $N_{\rm{exp}}$ fixed at 4 (right). $p$-value is $\approx 1$ for every fit shown except $N_{\rm{exp}}=1$.
      For the central fit we chose $N_{\rm{exp}}=4$ and $t_{\rm{min}}=4$.}
   \label{picorrplots}}
\end{figure}

We fit the two-point correlators using exponential forms with $N_{\rm{exp}}$ odd and $N_{\rm{exp}}$ even parity states,
increasing $N_{\rm{exp}}$ until the fit becomes stable.
Fit windows are chosen with $t_{\rm{min}}$ as small as possible while still giving both a good $p$-value and a consistent fit result.
For the non-zero momentum correlators the statistical errors grow quickly and we set $t_{\rm{max}}$ to the last time slice where the relative error on the correlation function is less than $30\%$.
Bayesian priors are employed to constrain excited-state contributions.
Fig.~\ref{picorrplots} displays some plots showing the stability of the fit results for the non-zero-momentum pion with respect to variations in the number of exponentials and the choice of $t_{\rm{min}}$. 
These results are similar to those on other ensembles and in the kaon cases.
These plots show that the fits are generally stable for $N_{\rm{exp}} \ge 3$ or $4$ in the range of $t_{\rm{min}}$ values we are interested in.
The fit values are also consistent for all $t_{\rm{min}}$ values in a reasonable range.

\begin{figure}
  {\centering \vspace{-8pt}
      \includegraphics[width=0.95\linewidth]{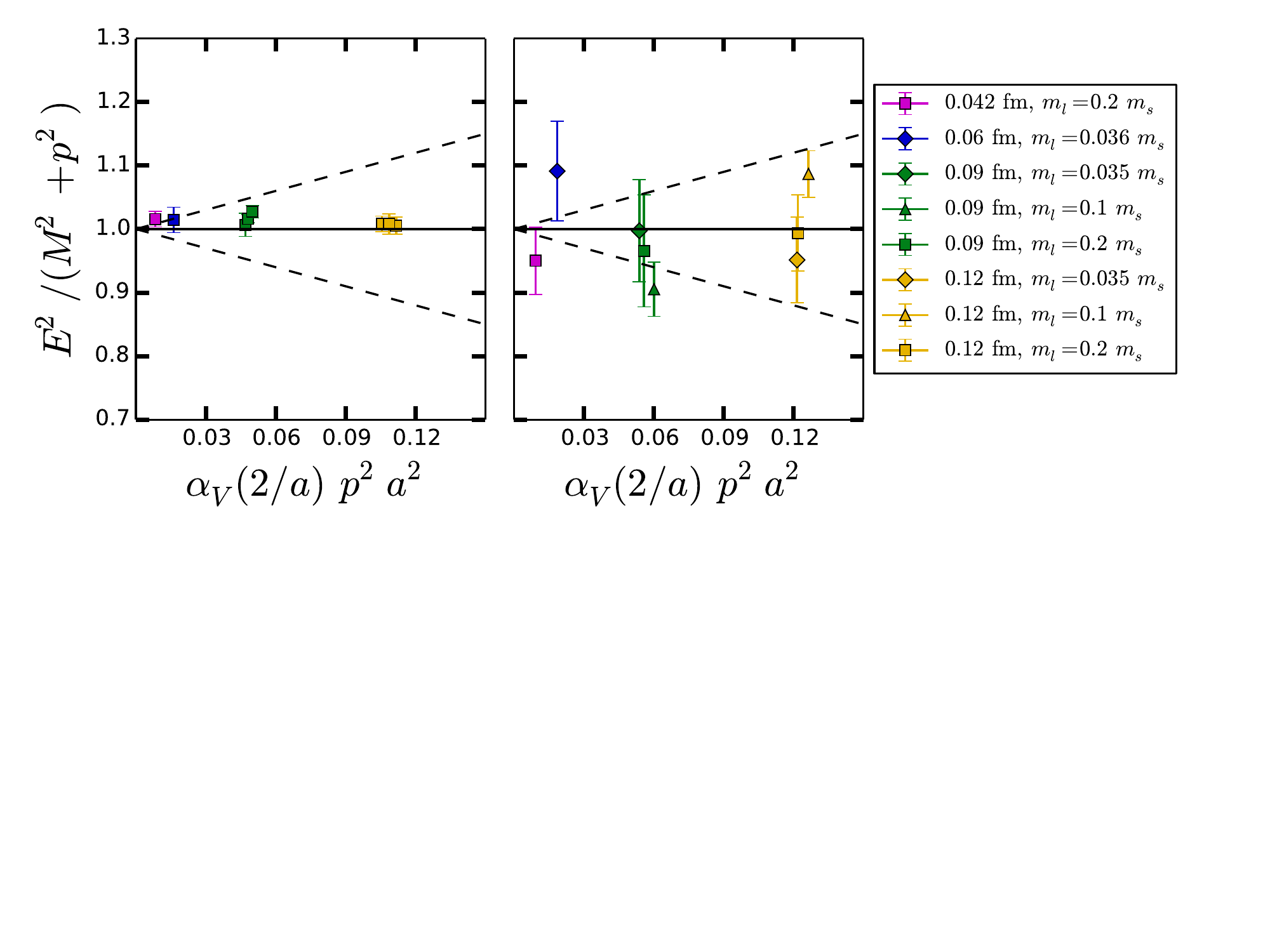} \\ \vspace{-136pt}
    \caption{Dispersion relation violations for the kaon (left) and pion (right) on each ensemble as a function of $\alpha_V (pa)^2$.
    The dashed lines show $1 \pm \alpha_V (pa)^2$ as guidelines for the expected scaling of the violations.}
   \label{dispersion}}
\end{figure}

The momentum transfer from the $D$ meson to the kaon (pion) in the $D$-meson rest frame is
\begin{equation}
  q^2 = M_{K(\pi)}^2+M_D^2 - 2M_DE_{K(\pi)},
\end{equation}
and the required momenta to achieve the value of $E_{K(\pi)}$ that results in $q^2=0$ is determined via the dispersion relation.
On the lattice the dispersion relation is not exact, with violations expected to scale as $\alpha_s (pa)^2$.
Figure~\ref{dispersion} shows the dispersion relation violations for the kaon and pion energies in our study.
The observed violations are within expectations; however, the statistical errors in the fitted energies seem to be the more significant cause of
the deviations from the dispersion relation.

\begin{figure}[t]
  {\centering \vspace{-18pt}
    \includegraphics[width=0.42\linewidth]{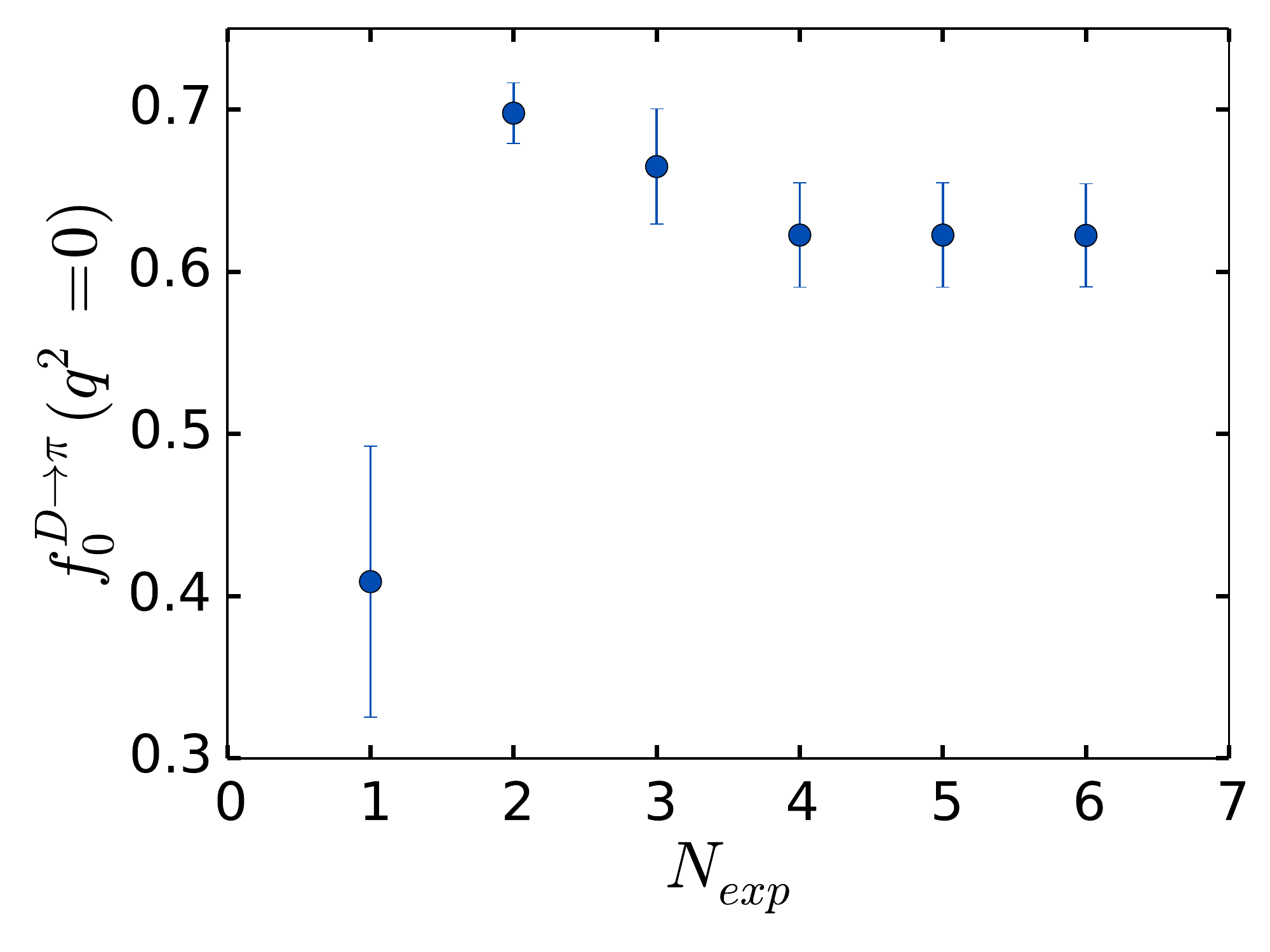}
    \includegraphics[width=0.42\linewidth]{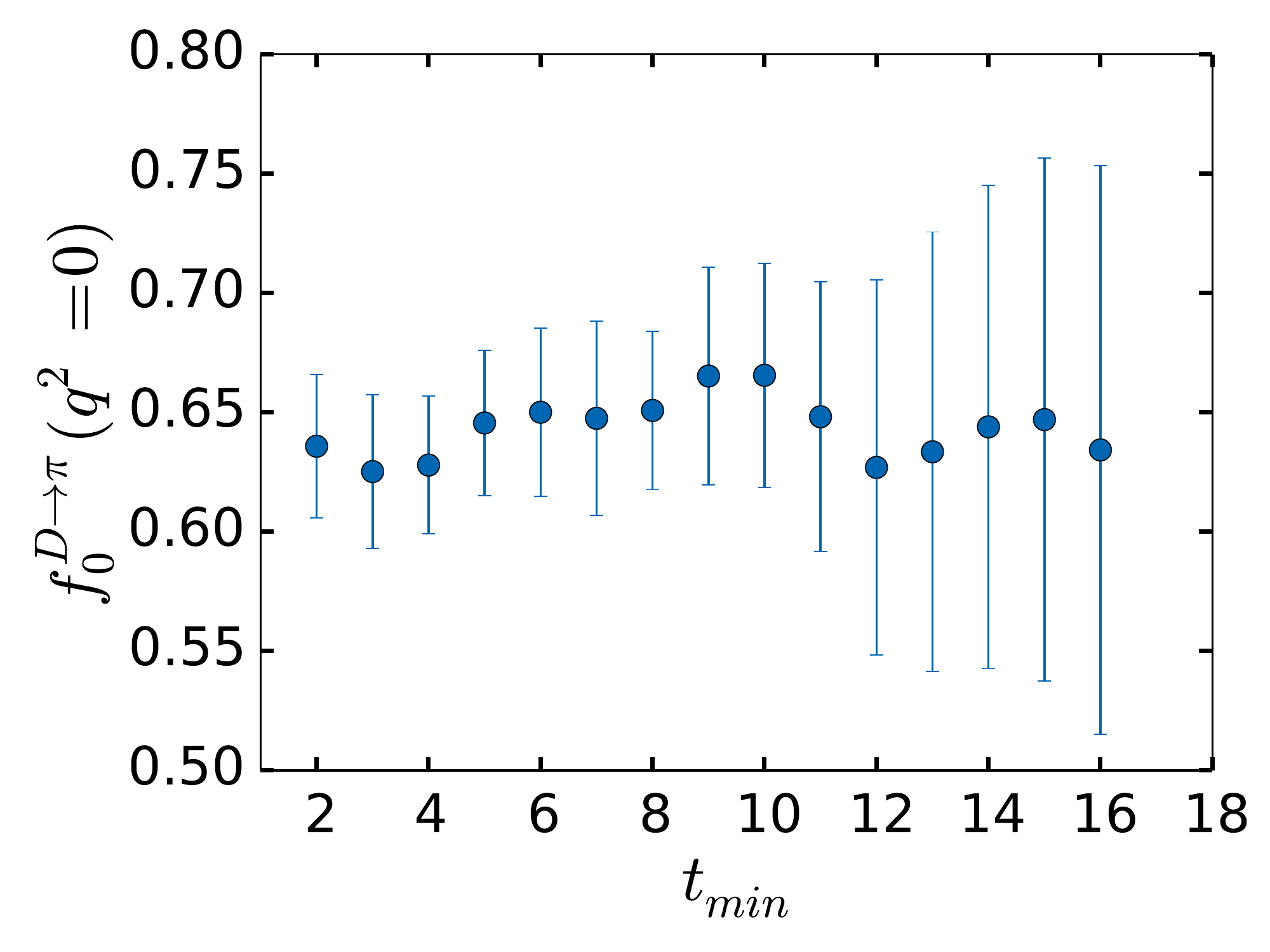} \\
    \caption{$D \to \pi$ form factor fit values from the 0.09 fm physical quark mass ensemble as a function of the number of states $N_{\rm{exp}}$ with $t_{\rm{min}}$ fixed at 4 (left)
      and as a function of $t_{\rm{min}}$ with $N_{\rm{exp}}$ fixed at 4 (right). $p$-value is $\approx 1$ for every fit shown except $N_{\rm{exp}} < 4$.
      For the central fit we chose $N_{\rm{exp}}=4$ and $t_{\rm{min}}=4$.}
   \label{DPTplots}}
\end{figure}

In order to fit our three-point correlators we carry out both simultaneous fits of the two-point and three-point functions
and sequential fits in which the results of the two-point fits are fed into the three-point fits as priors.
The two methods yield consistent results, but the sequential fits give slightly better stability.
Fit windows for the three-point correlators are chosen with $t_{\rm{min}}^{3pt} = t_{\rm{min}}^{K(\pi)}$ and $t_{\rm{max}}^{3pt} = T-t_{\rm{min}}^D$.
Only three out of five available $T$ choices are included in each fit, because including more than three
does not improve the errors or stability but sometimes causes the fit to not converge.
The $D\to K$ three-point fits have only slightly smaller statistical errors than the $D\to\pi$ fits,
while the kaon non-zero momentum two-point fits have significantly smaller statistical errors than the pion ones.
Fig.~\ref{DPTplots} shows plots of the form factor fit stability for $D\to \pi$, with the $D\to K$ fits being very similar.

\section{Chiral-continuum extrapolation}

\begin{figure} \vspace{-18pt}
    \includegraphics[width=0.60\linewidth]{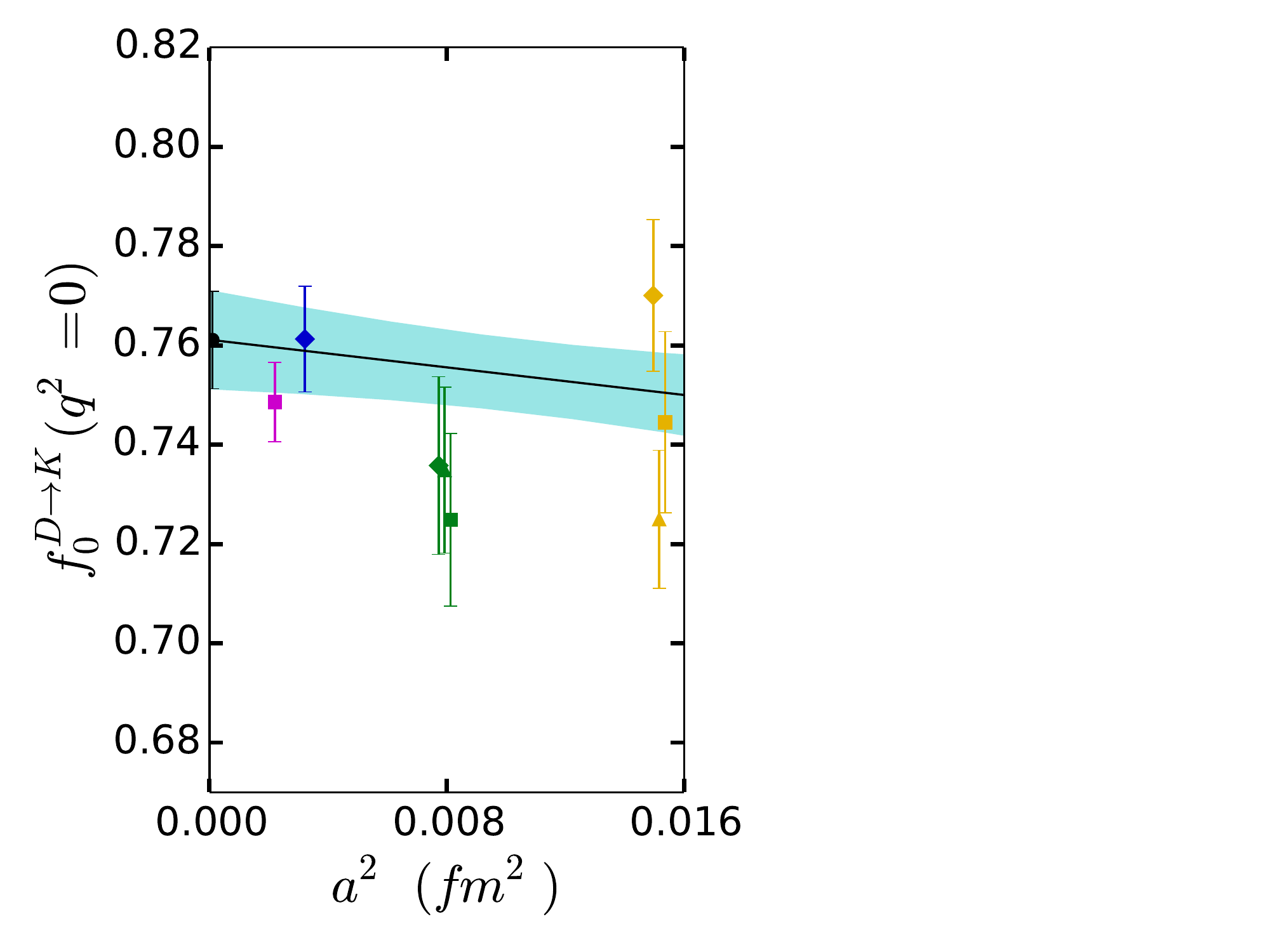}\hspace{-116pt}
    \includegraphics[width=0.60\linewidth]{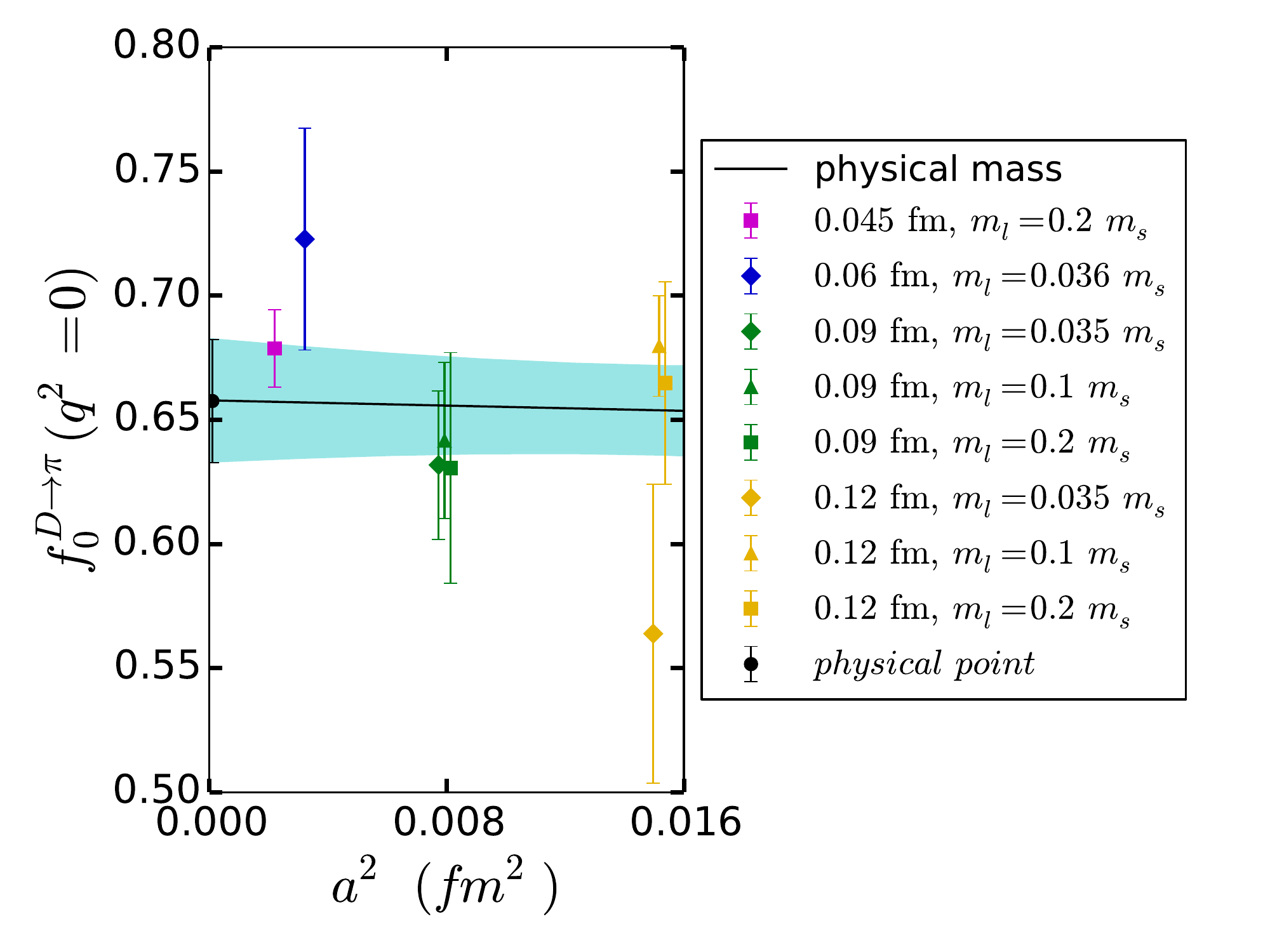}
    \caption{Chiral/continuum fits of $f_0^{D\to K}(0)$ (left) and $f_0^{D\to\pi}(0)$ (right) as a function of $m_l/m_s$.}\label{massratio}
  \end{figure}

We extrapolate the form factors to the physical light-quark mass and continuum using heavy-light-meson chiral perturbation theory (HM$\chi$PT).  
We employ the continuum next-to-leading-order chiral logarithms from Ref.~\cite{Becirevic} modified for energetic (``hard") pions and kaons in Ref.~\cite{HardPion}, 
and include analytic terms in the light- and strange-quark masses, kaon(pion) energies, and lattice spacing.  
Our central fit function has the simple form
\begin{eqnarray}
  f_0(q^2) &=& {c_0} ( 1 + df_{\rm{logs}}) + {c_1} \chi_\ell + {c_2} \chi_{a^2},
\end{eqnarray}
where $df_{\rm{logs}}$ are the chiral logarithms, $\chi_\ell$ and $\chi_{a^2}$ are analytic terms proportional to the light-quark mass and squared lattice spacing, and the coefficients $c_i$ are fit parameters.  
We construct dimensionless parameters $\chi_i$ such that the coefficients are expected to be of order one, and use priors $c_i = 0 \pm 2$.  
The chiral logarithms depend upon the $D^*$-$D$-$\pi$ coupling, which we constrain with a prior $g_\pi = 0.52 \pm 0.07$ to cover the spread of recent determinations~\cite{Becirevic:2012pf,Can:2012tx,Lees:2013uxa}.
The chiral-continuum extrapolations of the $D\to K$ and $D\to\pi$ form factors from our central fit function are shown in Fig.~\ref{massratio}

\begin{figure}[t]{\centering \vspace{0pt}
     ${\vcenter{\hbox{\includegraphics[width=0.45\linewidth]{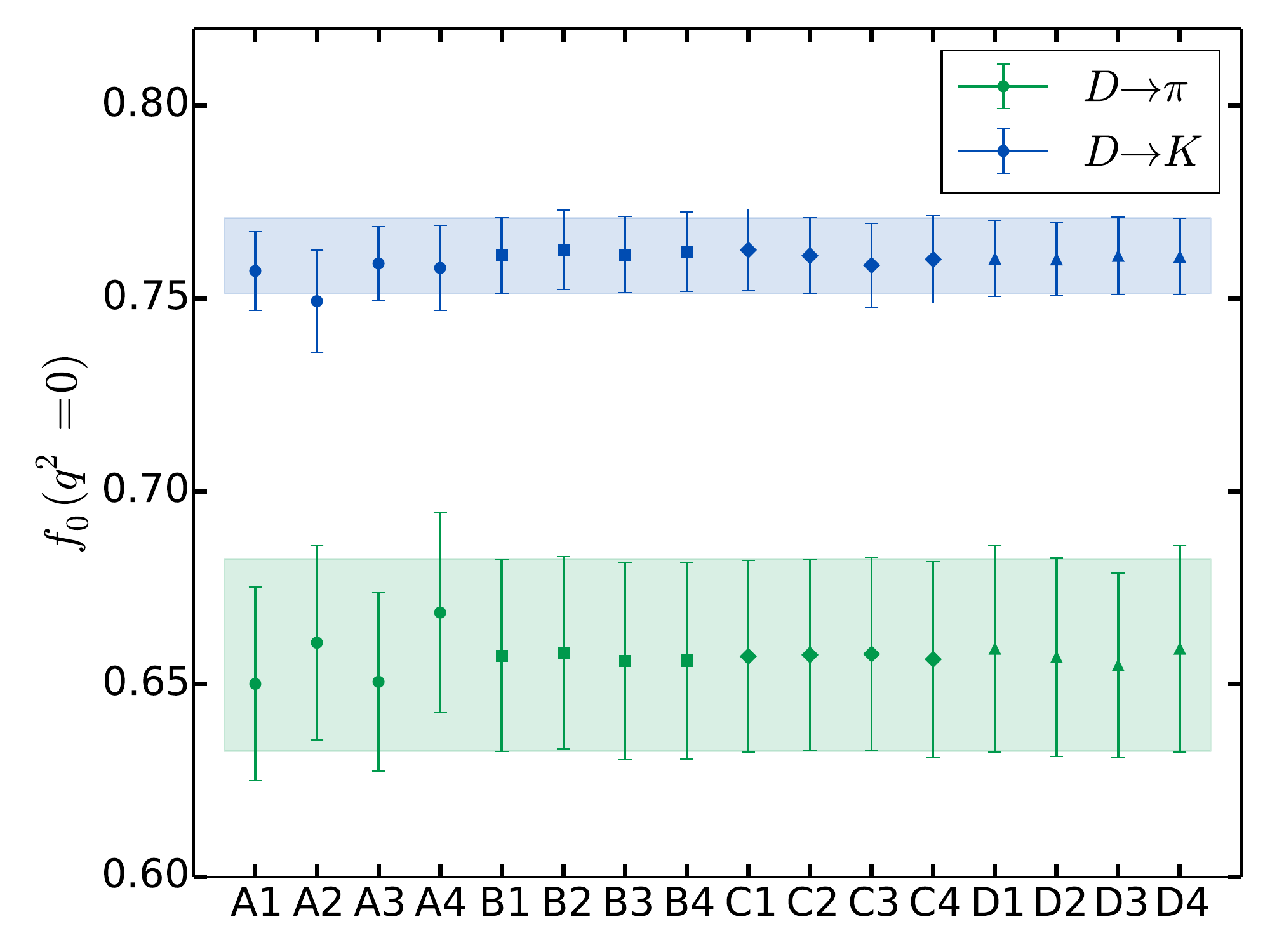}}}}$ \hspace{6pt}
     \caption{Stability under fit variations.
       ``A'' points are different fit window approaches in the correlator fits.
       ``B'' points include analytic NLO terms in the strange quark mass, sea quark masses and kaon (pion) energy.
       ``C'' points add NNLO terms in combinations of the light quark mass and the square of the lattice spacing.
       ``D'' points are from different ways of parameterizing the lattice spacing dependence.
     }
     \label{stability}}
 \end{figure}

We do not include analytic terms in the strange-quark mass, kaon (pion) energy, or sea-quark mass in our central fit
because the first two are approximately constant across all of our ensembles, and the sea quark mass is either the same as the valence
quark mass or differs by only a small mistuning.
Therefore, we cannot resolve these dependencies, but expect the corresponding errors to be small since our parameters are chosen very close to their physical values.
As illustrated in Fig.~\ref{stability} we do consider these and other chiral-continuum fit variations in our systematic error analysis and find the fits stable under the inclusion of such terms.

\section{Conclusions}

Based on the fit stability shown in Fig.~\ref{stability}, we construct a preliminary error budget as shown in Table~1.  
The chiral fit error includes statistical errors as well as those from discretization effects and truncation of the chiral expansion.
In order to estimate the size of sea-quark mistuning effects, we repeat the chiral fit with the valence-quark masses changed to the sea-quark masses, 
and take the difference between the two results as the error denoted ``$m_s^{\rm{val}} \ne m_s^{\rm{sea}}$".
To estimate the error from the lattice-spacing uncertainty, we repeated the chiral fit twice for each lattice spacing, 
with the value of the lattice spacing replaced by first $a+\sigma$ and then $a-\sigma$ from the mass-independent scale setting \cite{FNALMILC_DC}. 
We take the largest observed difference as the scale error.
The listed finite-volume error is taken from our recent $K \to \pi$ calculation on the MILC HISQ configurations~\cite{Bazavov:2013maa}, which included extra ensembles at different volumes.
We intend to estimate this error directly by performing calculations on additional volumes.

In these proceedings we summarize our progress on calculating the $D\to K$ and $D\to \pi$ semileptonic form factors 
directly at $q^2=0$.
To complete our analysis, we are including an additional $0.06$ fm ensemble to help resolve the lattice-spacing dependence.
In addition, we plan to analyze ensembles with different spatial volumes to assess directly the size of finite-volume effects.
We will also employ $\chi$PT expressions that incorporate taste-breaking discretization effects from staggered quarks and otherwise refine our error analysis.
Upon completion of this project, we anticipate total errors of approximately $2\%$ and $5\%$ for $D\to K$ and $D\to \pi$, respectively.
In the future,  we plan to undertake a calculation 
of $f_0(q^2)$ and $f_+(q^2)$ at a variety of $q^2$ values to obtain both the normalization and shape of the form factors. 
%We will employ a $z$-expansion to parameterize the form factors over the full kinematic range,
%and to combine them with experimental measurements from the whole $q^2$ range to improve the determinations of the CKM matrix elements $\lvert{V_{cs}}\rvert$ and $\lvert{V_{cd}}\rvert$.

\begin{table}\centering \vspace{-18pt}
  {\fontsize{12pt}{7.2}\selectfont
  \begin{tabular}{l d d}
    \hline \hline 
    Source of  & \multicolumn{2}{c}{\% Error} \\
    uncertainty &  \multicolumn{1}{c}{$f^{D\to\pi}_+(0)$} & \multicolumn{1}{c}{$f^{D\to K}_+(0)$}  \\ \hline
    Statistics $\oplus$ $\chi$PT $\oplus$ $a^2$ $\oplus$ $g_\pi$           & 4.5    & 1.5  \\
    $m_s^{val} \ne m_s^{sea}$    & 0.04    & 0.15 \\
    Finite volume & (0.2)    & (0.2)  \\
    Scale $a$                 & 0.02   & 0.3 \\
    \hline
    Total                     & 4.5    & 1.6 \\ \hline \hline
  \end{tabular}
  }
  \caption{Preliminary error budgets for $f_+(0)$ for $D\to K$ and $D\to\pi$.
      The first error includes statistical errors from the simulation and systematics associated with the chiral-continuum fit.
      The finite-volume error is taken from our calculation of the $K\to\pi$ form factor on the same ensembles~\cite{Bazavov:2013maa}.
      }\label{errors}
\end{table}

\acknowledgments

Computations for this work were carried out with resources provided by
the USQCD Collaboration, the ALCF, and NERSC, which are funded by the
U.S. Department of Energy (DOE); and with resources provided by NCAR,
NCSA, NICS, TACC, and Blue Waters which are funded through the
U.S. National Science Foundation (NSF).
Fermilab is operated by Fermi Research Alliance, LLC, under Contract No. DE-AC02-07CH11359 with the U.S. Department of Energy.
Authors of this work were
also supported in part through individual grants by the DOE and NSF
(U.S); by MICINN, the Ram\'on y Cajal program, and the Junta de
Andaluc\'ia (Spain); by the European Commission; and by the German Excellence Initiative.


\begin{thebibliography}{99}

\bibitem{FNALMILC_DC}
  A.~Bazavov {\it et al.} [Fermilab Lattice and MILC],
  Phys.\ Rev.\ D {\bf 90}, 074509 (2014),
  [arXiv:1407.3772 [hep-lat]].
\bibitem{HPQCD_1}
  H.~Na  {\it et al.} [HPQCD], Phys.\ Rev.\ D {\bf 84}, 114505 (2011), [arXiv:1109.1501 [hep-lat]].
\bibitem{HPQCD_2}
  J.~Koponen {\it et al.} [HPQCD], arXiv:1305.1462 [hep-lat].
\bibitem{HFAG}
  Y.~Amhis {\it et al.} [Heavy Flavor Averaging Group], arXiv:1412.7515 [hep-ex].
\bibitem{Rosner:2015wva}
  J.~L.~Rosner, S.~Stone and R.~S.~Van de Water,
  arXiv:1509.02220 [hep-ph].
\bibitem{HPQCD_scalar}
  H.~Na {\it et al.} [HPQCD], Phys.\ Rev.\ D {\bf 82}, 114506 (2010)
  [arXiv:1008.4562 [hep-lat]].

%
\bibitem{Bazavov:2010ru}
  A.~Bazavov {\it et al.} [MILC],
  %``Scaling studies of QCD with the dynamical HISQ action,''
  Phys.\ Rev.\ D {\bf 82}, 074501 (2010) [arXiv:1004.0342];
  Phys.\ Rev.\ D {\bf 87}, 054505 (2013) [arXiv:1212.4768].
  %%CITATION = ARXIV:1212.4768;%%                                                                               
%
\bibitem{Follana:2006rc}
  E.~Follana {\it et al.} [HPQCD],
  %``Highly improved staggered quarks on the lattice, with applications to charm physics,''
  Phys.\ Rev.\ D {\bf 75}, 054502 (2007)
  [hep-lat/0610092].
  %%CITATION = HEP-LAT/0610092;%%
%
\bibitem{milc_hisq}
%\cite{Bazavov:2009jc}
  A.~Bazavov {\it et al.}  [MILC],
  %``HISQ action in dynamical simulations,''
  PoS LATTICE2008, 033 (2008)
  [arXiv:0903.0874];
  %%CITATION = POSCI,LATTICE2008,033;%%
  % A.~Bazavov {\it et al.},
  %``Progress on four flavor QCD with the HISQ action,''
  PoS LAT2009 123 (2009)
  [arXiv:0911.0869];
  % Lattice 2010 (Sardinia)
  % Simulations with dynamical HISQ quarks,
  PoS(Lattice 2010), 320 (2010)
  [arXiv:1012.1265].
%
\bibitem{Becirevic}
  D.~Becirevic, S.~Prelovsek and J.~Zupan,
  Phys.\ Rev.\ D {\bf 68}, 074003 (2003)
  [hep-lat/0305001].
\bibitem{HardPion}
  J.~Bijnens and I.~Jemos,  Nucl.\ Phys.\ B {\bf 840}, 54 (2010); (erratum) Nucl.\ Phys.\ B {\bf 844}, 182 (2011)
  [arXiv:1006.1197 [hep-ph]].
%\cite{Becirevic:2012pf}

\bibitem{Becirevic:2012pf} 
  D.~Becirevic and F.~Sanfilippo,
  %``Theoretical estimate of the $D^* \to D\pi$ decay rate,''
  Phys.\ Lett.\ B {\bf 721}, 94 (2013)
  [arXiv:1210.5410 [hep-lat]].
%\cite{Lees:2013uxa}
%\cite{Can:2012tx}
\bibitem{Can:2012tx} 
  K.~U.~Can, G.~Erkol, M.~Oka, A.~Ozpineci and T.~T.~Takahashi,
  %``Vector and axial-vector couplings of D and D* mesons in 2+1 flavor Lattice QCD,''
  Phys.\ Lett.\ B {\bf 719}, 103 (2013)
  [arXiv:1210.0869 [hep-lat]].
\bibitem{Lees:2013uxa} 
 J.~P.~Lees {\it et al.} [BaBar],
  %``Measurement of the $D^*(2010)^+$ natural line width and the $D^*(2010)^+ - D^0$ mass difference,''
  Phys.\ Rev.\ D {\bf 88}, 052003 (2013);
  (erratum) Phys.\ Rev.\ D {\bf 88} 079902 (2013)
  [arXiv:1304.5009 [hep-ex]].

\bibitem{Bazavov:2013maa} 
  A.~Bazavov {\it et al.} [Fermilab Lattice and MILC],
  %``Determination of $|V_{us}|$ from a lattice-QCD calculation of the $K\to\pi\ell\nu$ semileptonic form factor with physical quark masses,''
  Phys.\ Rev.\ Lett.\  {\bf 112}, 112001 (2014)
  [arXiv:1312.1228 [hep-ph]].



\end{thebibliography}
\end{document}